\documentclass{emulateapj}
\usepackage{url}
\def\arcsec{$^{\prime\prime}$}
\newcommand\msun{\rm{M}_\odot}
\newcommand\kms{km s$^{-1}$}

\newcommand\ha{$\rm{H}\alpha$}
\newcommand\hb{$\rm{H}\beta$}
\def\oiii{[O\,{\sc iii}]}
\def\oii{[O\,{\sc ii}]}

\shorttitle{IFU Star Formation Histories}
\shortauthors{Yoachim et al.}

\begin{document}

\title{IFU Spectroscopy of the Stellar Disk Truncation Region of NGC 6155\footnote{ This paper includes data taken at The McDonald Observatory of The University of Texas at Austin.}}

\author{Peter Yoachim\altaffilmark{1}, Rok Ro{\v s}kar\altaffilmark{2}, Victor P. Debattista\altaffilmark{3}
} \altaffiltext{1}{Department of
Astronomy and McDonald Observatory, University of Texas, Austin, TX
78712; {yoachim@astro.as.utexas.edu}}
\altaffiltext{2}{Department of Astronomy, University of Washington, Box 351580, Seattle WA, 98195}
\altaffiltext{3}{RCUK Fellow, Jeremiah Horrocks Institute, University of Central Lancashire, Preston PR1 2HE}

\begin{abstract}

Like the majority of spiral galaxies, NGC 6155 exhibits an exponential surface brightness profile that steepens significantly at large radii.  Using the VIRUS-P IFU spectrograph, we have gathered spatially resolved spectra of the system.  Modifying the GANDALF spectral fitting routine for use on the complex stellar populations found in spirals, we find that the average stellar ages increase significantly beyond the profile break radius.  This result is in good agreement with recent simulations that predict the outskirts of disk galaxies are populated through stellar migration.  With the ability to bin multiple fibers, we are able to measure stellar population ages down to $\mu_V\sim24$ mag/sq arcsec.

\end{abstract}

\section{Introduction}

The nature of how and where stellar disks end can shed light on the limits of star formation and the importance of secular evolution.  Stellar disks tend to be embedded in much larger HI disks \citep{Bosma81,Broeils97,Begum05}, showing that while there are baryons in the outskirts of galaxies, they are not being converted into stars efficiently.  It is often assumed that there is a surface density threshold for star formation \citep{Kennicutt89} which would provide a natural mechanism for truncating the stellar component of galaxies.

The surface brightness profiles of spiral galaxies are commonly fit with exponentials \citep{Freeman70}.  While the exponential function is a convenient description, few astronomers would suggest that stellar disks extend to infinite radii.  Observations of edge-on systems originally suggested that stellar disks have well defined truncation radii \citep{vdk81a,Kregel04}.  Using large samples of nearly face-on galaxies, \citet{Pohlen06} found a variety of behaviors at large galactic radii.  \citet{Pohlen06} find that only 10\% of disks are well fit by a single exponential light profile, while in 60\% of their sample the surface brightness profile is well fit by a downbending broken exponental, with a break radius 1.5-4.5 times the inner scale length.  The final 30\% of their sample are described with a broken upbending exponential profile.  Unlike the edge-on studies, they do not find evidence for sharp truncations.

\citet{Roskar08} present a possible formation mechanism for surface brightness profile breaks.  They find that a star formation threshold seeds the onset of a truncation, but radial migration populates the disk beyond the break producing a down-bending broken profile.  Since the stellar migration is a random walk, only ancient stars have had enough time to get to very large radii beyond where star formation shuts off.  These simulations thus predict that the break radius should correspond with an increase in the average stellar ages.  The formation and appearance of outer disks has since been investigated in cosmological simulations by several groups who also find an upturn in age at the break radius \citep{Sanchez09b,Martinez09}. However, the relative importance of radial migration and in-situ star formation in a fully-cosmological setting remains an open question.

While simulations find that profile breaks can be caused by a combination of a star formation threshold and stellar migration, observations suggest that star formation can proceed even in the outskirts of disks.  Deep \ha\ imaging by \citet{Ferguson98} show signs of extended star formation.  \citet{Thilker08} also find that $\sim30$\% of spiral galaxies have UV emission extending beyond their traditional star formation threshold radius.  Similarly, \citet{Herbertfort09} find clustering of point sources around NGC 3184, implying star cluster formation in the outer disk.

Recent observations of resolved stars have found a variety of results for stellar truncation regions.  \citet{Williams09} find that in low-mass M33 the the age gradient changes from negative to positive at the truncation radius.  Meanwhile, \citet{deJong07} find that the truncation region in NGC 4244 is the same for stars of all ages.  \citet{deJong07} interpret this as a sign the truncation is the result of a dynamical interaction, but these results were subsequently shown to be consistent with the radial migration model \citep{Roskar08}.  See \citet{Vlajic10} for a review of the recent advances in studying the outskirts of spiral disks.  The observational results point to an interesting conundrum that while many studies find active star formation in the outskirts of disk galaxies, the stars in the region are often older than the inner disk.

To date, truncation regions have been studied with broadband colors and resolved stars.  There are a limited number of systems we can resolve, and faint broadband surface photometry is dominated by systematic errors associated with flat-fielding and sky subtraction.  While there have been searches targeting emission lines \citep{Ferg98,Chirst10}, in this letter we present the first spectroscopic study of stellar continuum spectral features across a surface brightness profile break.

\section{Observations}

\begin{figure*}[h]
\plottwo{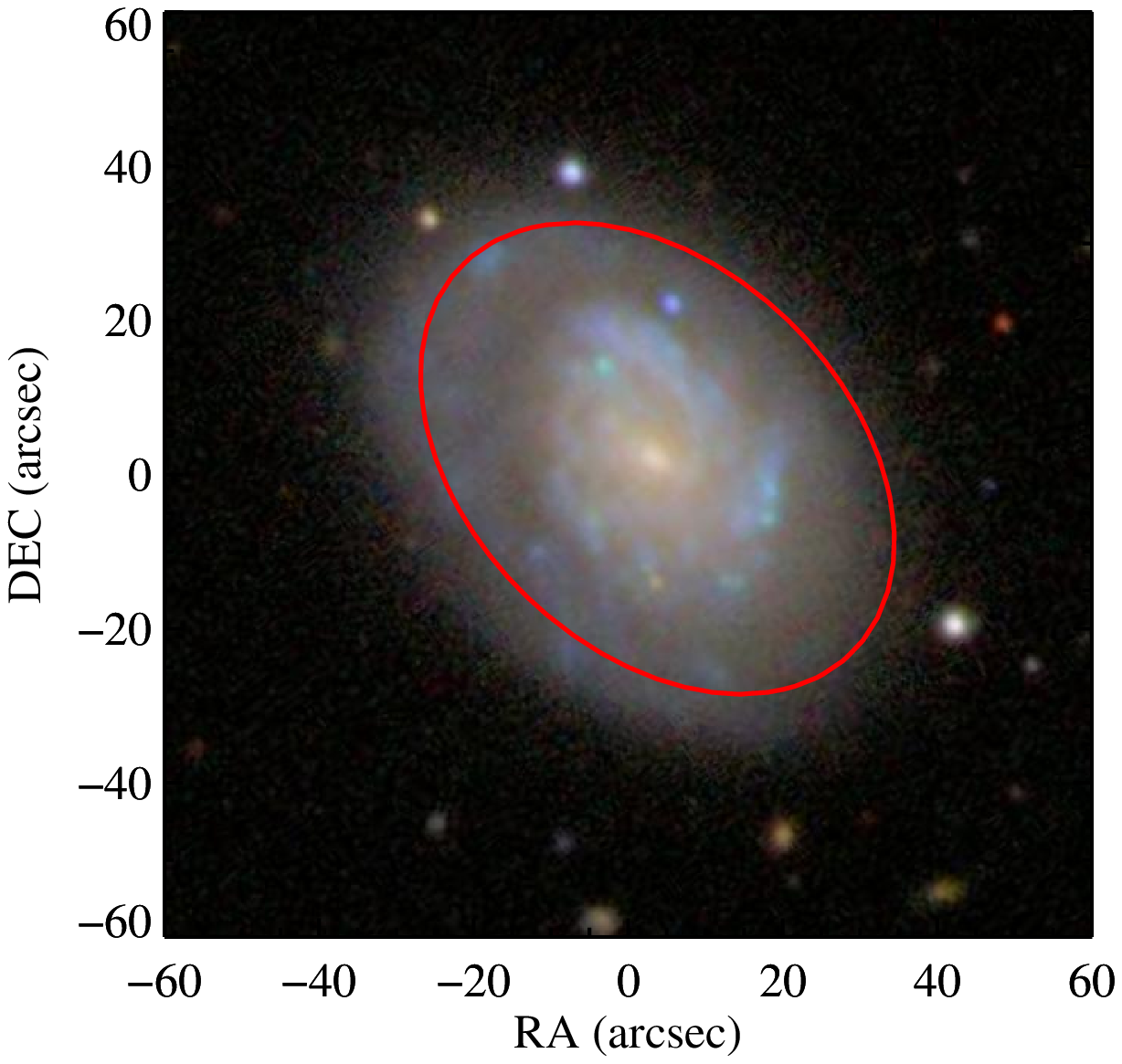}{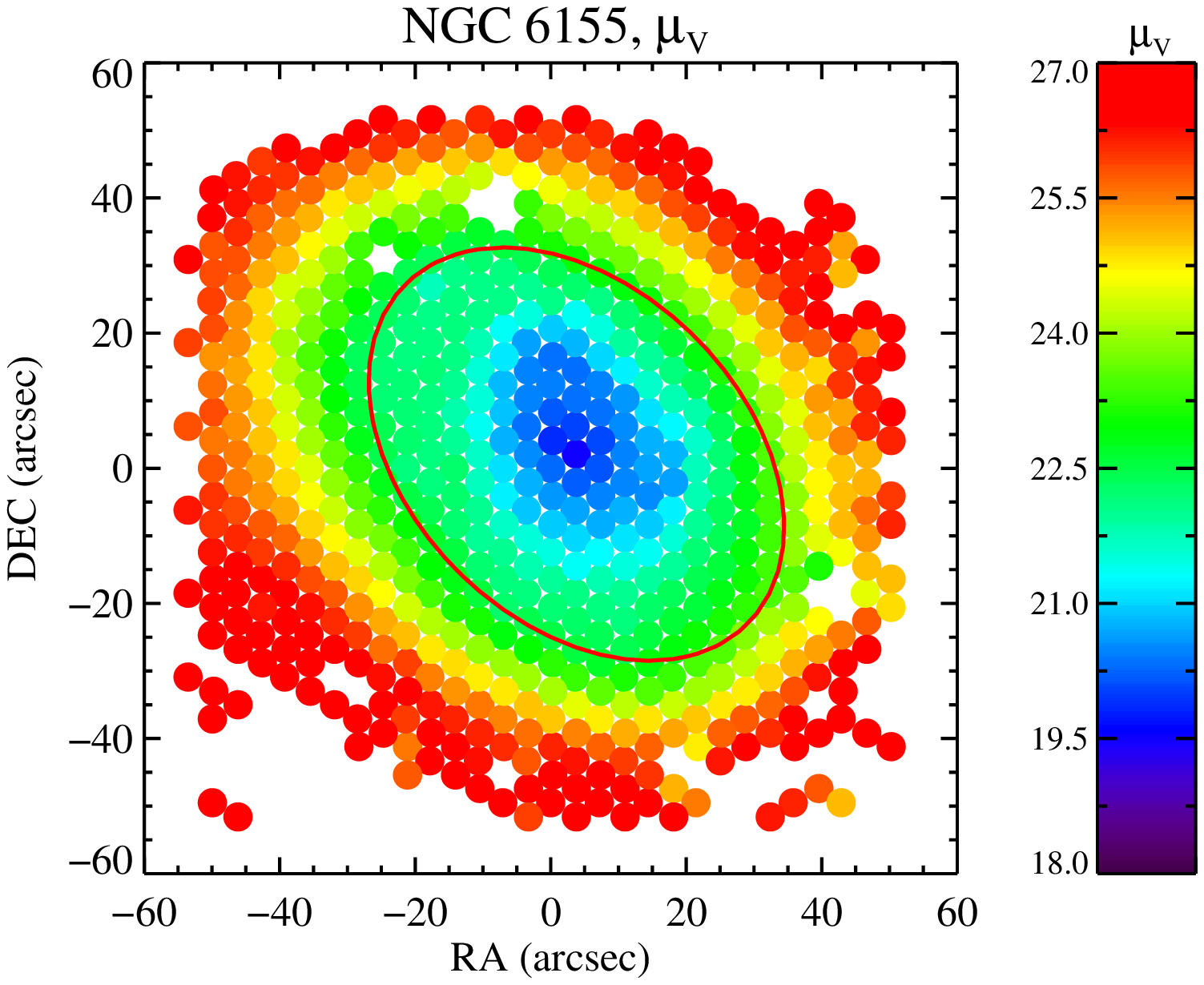}
\plottwo{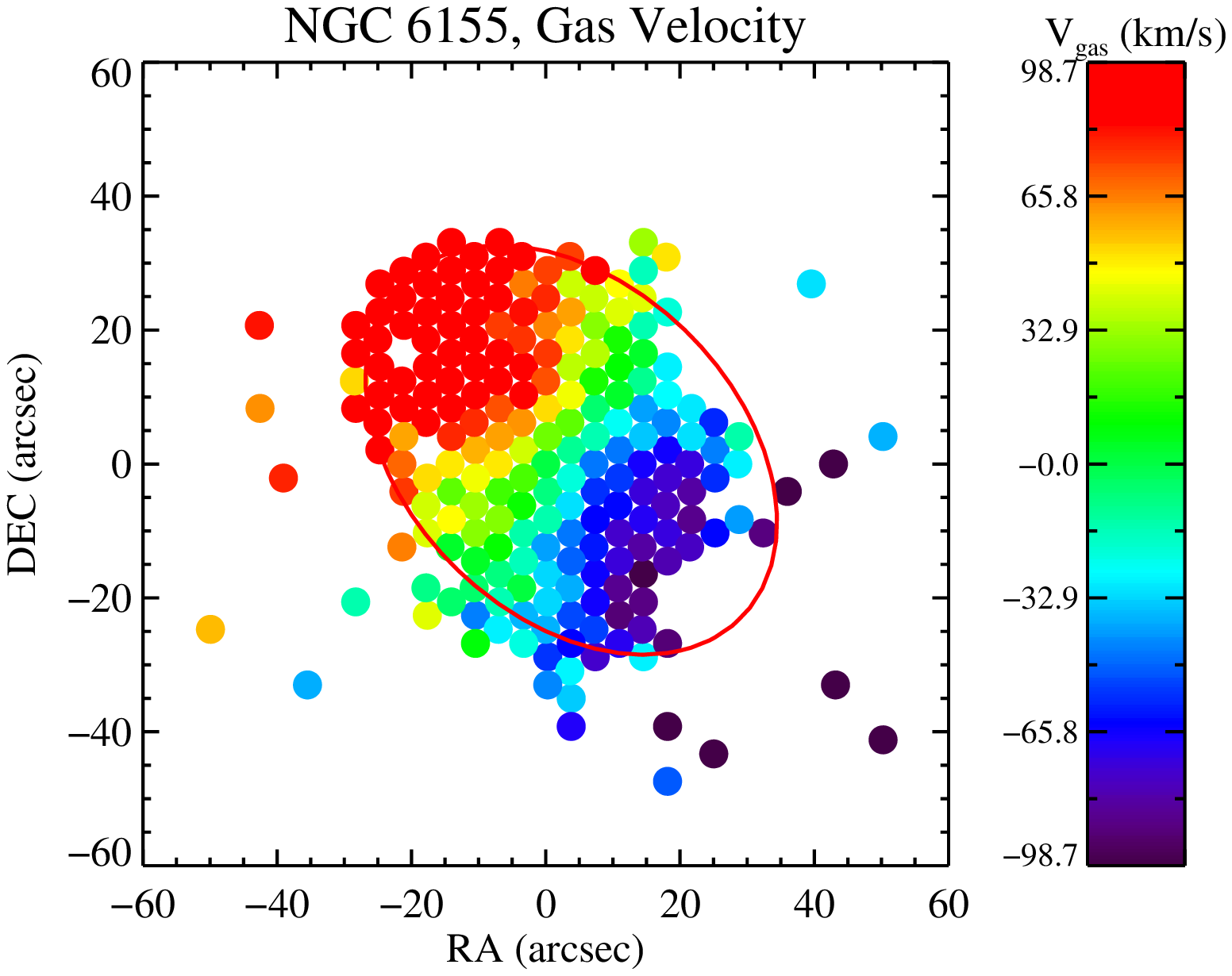}{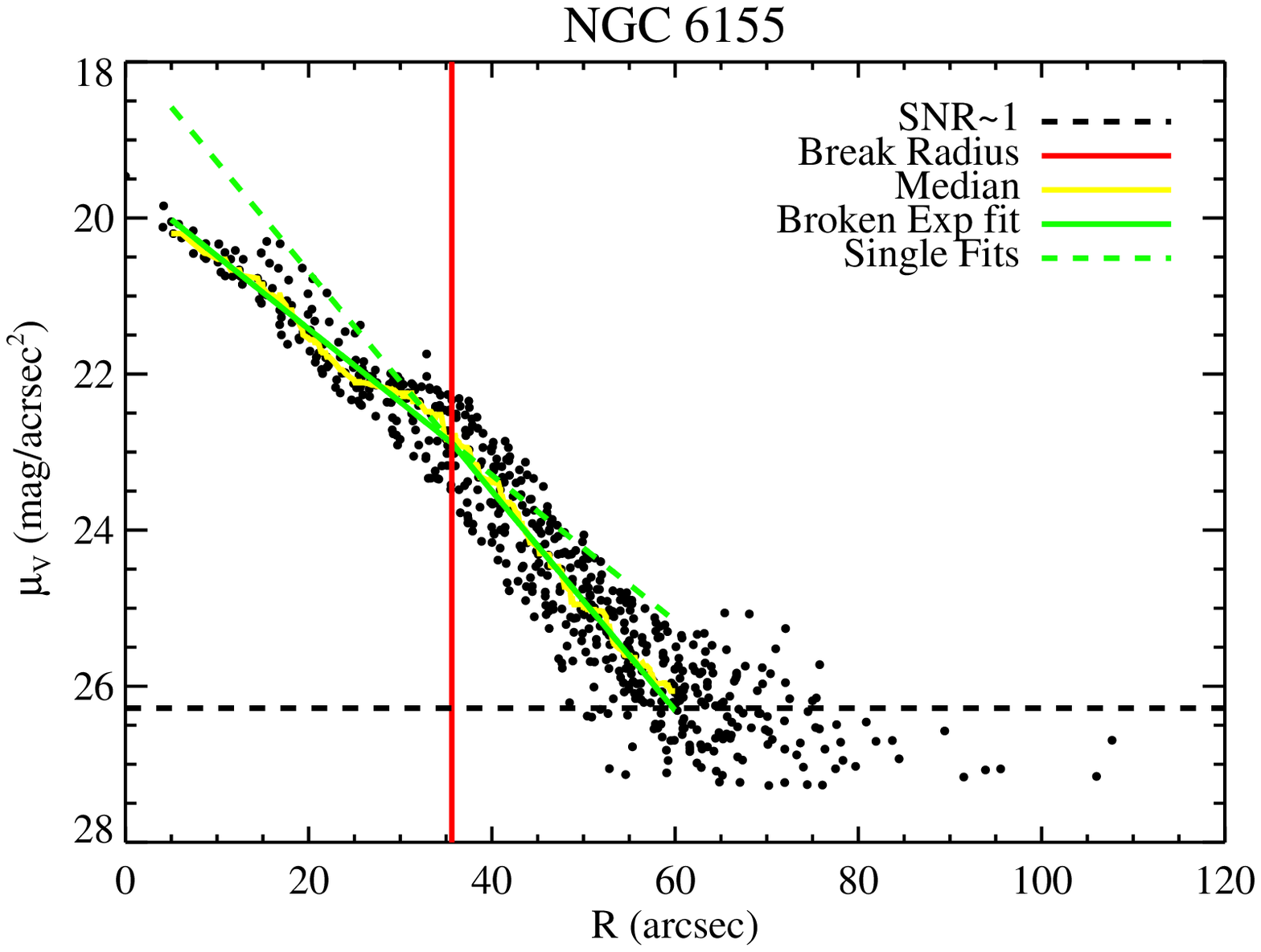}
\caption{Upper Left:  SDSS image of NGC 6155 (North is up and East is right).  Upper Right:  Surface brightness measured from VIRUS-P fiber spectra.  Points are the same size as the VIRUS-P fibers.  Fibers containing stars or bright background galaxies have been masked.  Lower Left:  Velocity field measured from emission lines.  Lower Right:  Surface brightness profile as measured by individual fibers.   A green line shows the best-fit broken exponential.  The dashed green line shows the result of extrapolating the inner and outer region fits.  A yellow curve shows a running median with an 5\arcsec\ window.  The dashed line shows the level where the SNR of a single fiber is unity.  In each panel, a red curve marks the best-fit break radius.\label{sb}}
\end{figure*}

Observations of NGC 6155 (RA 16:26:08, DEC +48:22:01, J2000) were taken with the Visible Integral field Replicable Unit Spectrograph Prototype (VIRUS-P) \citep{Hill08} on the McDonald Observatory 2.7m Harlan J. Smith telescope on April 9 and 11, 2008.  VIRUS-P is a 1.7\arcmin x 1.7\arcmin\ field-of-view IFU spectrograph with 246 4.3\arcsec\ diameter fibers.  Observations were taken in a three-step dither pattern to attain full spatial coverage.  We observed four 20 minute exposures at each dither position for a total of four hours of exposure time.  The data were reduced with the VACCINE package (Adams et al. \emph{in prep}) in combination with custom IDL routines.  Wavelength calibration was performed using observations of Hg and Cd arc lamps.  Twilight sky exposures were used to generate flat-field frames.  We observed the stars Feige 34 and BD+33 2642 to establish flux calibration.  The final result is a data cube with 738 spectra filling a 1.7\arcmin x 1.7\arcmin\ field of view and a FWHM resolution of 5.3 \AA\ with 2.2 \AA\ pixels and a wavelength range of 3550-5580 \AA.  We mask a 10 \AA\ region around the bright 5577 \AA\ O I skyline.

Using the $V$-band surface brightness measured from our data cube we find an inclination of 46 degrees and position angle of 135 degrees (East of North).  These values are consistent with the measured velocity field.  The radial surface brightness profile is well fit with a broken exponential with a break radius of 36\arcsec, where the scale length changes from 11.6\arcsec\ to 7.7\arcsec.  We fit the region of 5\arcsec$<$ R $<$60\arcsec\ to avoid any central bulge and outer regions of low SNR.  \citet{Pohlen06} fit the SDSS image of NGC 6155 and find a break radius of 34\arcsec\ and scale lengths of 12.2 (13.1) and 8.1 (7.3) in the $r$ ($g$) band.  Our values are an excellent match to the \citet{Pohlen06} fits, especially considering our lower spatial resolution.  Figure~\ref{sb} shows our fiber photometry, the velocity field measured from gas emission lines (\hb, \oiii, and \oii), and our radial profile fit.

We re-binned the spectra to a common velocity.  Spectra were then co-added in 4\arcsec\ elliptical annuli of constant galactic radius to reach adequate SNR at large radii.  Fibers that contained bright stars or background galaxies were masked.  Our final binned spectra have SNR/\AA$\sim$320 in the inner regions and $\sim45$ beyond the break radius.  We assume a Virgo infall corrected redshift to NGC 6155 of 2549 \kms (38.3 Mpc).

\section{Stellar Population Synthesis}

Historically, prominent individual spectral absorption features have been used to constrain the flux-weighted average age of stellar populations.  Systems based on stellar absorption features like Lick indices have proven useful for old elliptical stellar populations \citep[e.g.,][]{Worthey94,Trager00}, but become difficult to interpret when there are multiple stellar populations or extended star formation histories \citep{Serra07}.  

There are now a host of packages developed to measure star formation histories (SFHs) from integrated spectra \citep[e.g.,][]{Ocvirk06,MacArthur09,Tojeiro07b,Koleva09,Chilingarian07}.  All of these codes attempt to model observed galactic spectra as a linear combination of single-burst simple stellar populations (SSPs).  While it should be possible in theory to reconstruct a star formation history as a series of SSPs, in practice, we find that the problem becomes degenerate at low SNR.  There is also a strong degeneracy between including dust extinction and the presence of an old red stellar population.  In general, we find that it is very difficult to constrain an average stellar age using a linear combination of SSPs. 

Unlike spheroidal systems that can be well described as having only old stars, spiral disks overwhelmingly display a mix of very old and newly formed stars.  Observations of resolved stars in nearby spiral galaxies show that star formation histories are extended at all radii \citep{Williams09,Williams08,Gogarten10,Barker06}.  Similarly, the solar neighborhood in the Milky Way contains young stars, intermediate age stars like the sun, and very old thick disk stars.

Motivated by the observational results that even the outskirts of galaxies have a mix of stellar ages, we have constructed a large library of extended SFHs and find the {\emph{best fitting single spectrum}}.  We use the latest stellar synthesis package from Charlot \& Bruzual \citep{Charlot07,Bruzual07b} to generate a suite of composite stellar population spectra.  For all the models, we assume an exponential star formation history of the form
\begin{equation}
\psi(t,\mathrm{[Z/H]})=\rm{e}^{-t/\tau}\msun yr^{-1}
\end{equation}
where $\psi$ is the star formation rate and $\tau$ is the star forming time-scale.  We include both positive and negative values of $\tau$.  All the models are computed at an age of 13 Gyr.  We generate 1664 total model spectra, with values of $\tau$ ranging from -200 Gyr to -0.25 Gyr, and 0.5 Gyr to 200 Gyr.  Negative values of $\tau$ represent an increasing star formation rate.   Small absolute values of $\tau$ represent extreme stellar populations (dominated by very old stars or newly formed stars), while large absolute values of $\tau$ are effectively constant star formation histories.  Our models span mass-weighted ages of 15 Myr to 12.7 Gyr.  The metallicity values run from Z=10$^{-4}$ to 0.1 (with solar defined as Z=0.02).  We use the Padova 1994 stellar evolutionary tracks and a Chabrier initial mass function (IMF).  

We do not include any metallicity evolution in the models, thus our fits only return a flux-weighted average metallicity.  The advantage of our method is that it runs fairly quickly because of the limited parameter space and the SFHs are motivated by observations of spatially resolved stellar populations.  The disadvantage is that our metallicity measure is fairly crude since we do not include chemical evolution.  Also, we can not expect our fits to be robust if a galaxy has undergone strong episodic bursts of star formation.  We do not include $\alpha$-element variations in our spectral models, however as a late-type spiral, we would not expect NGC 6155 to be particularly $\alpha-$enhanced.  We emphasize that our fitted SFHs are a measure of the current stellar populations, after radial mixing and azimuthal binning, and not a measure of the actual star formation history at a fixed radius.

\begin{figure*}[h]
\epsscale{.55}
\plotone{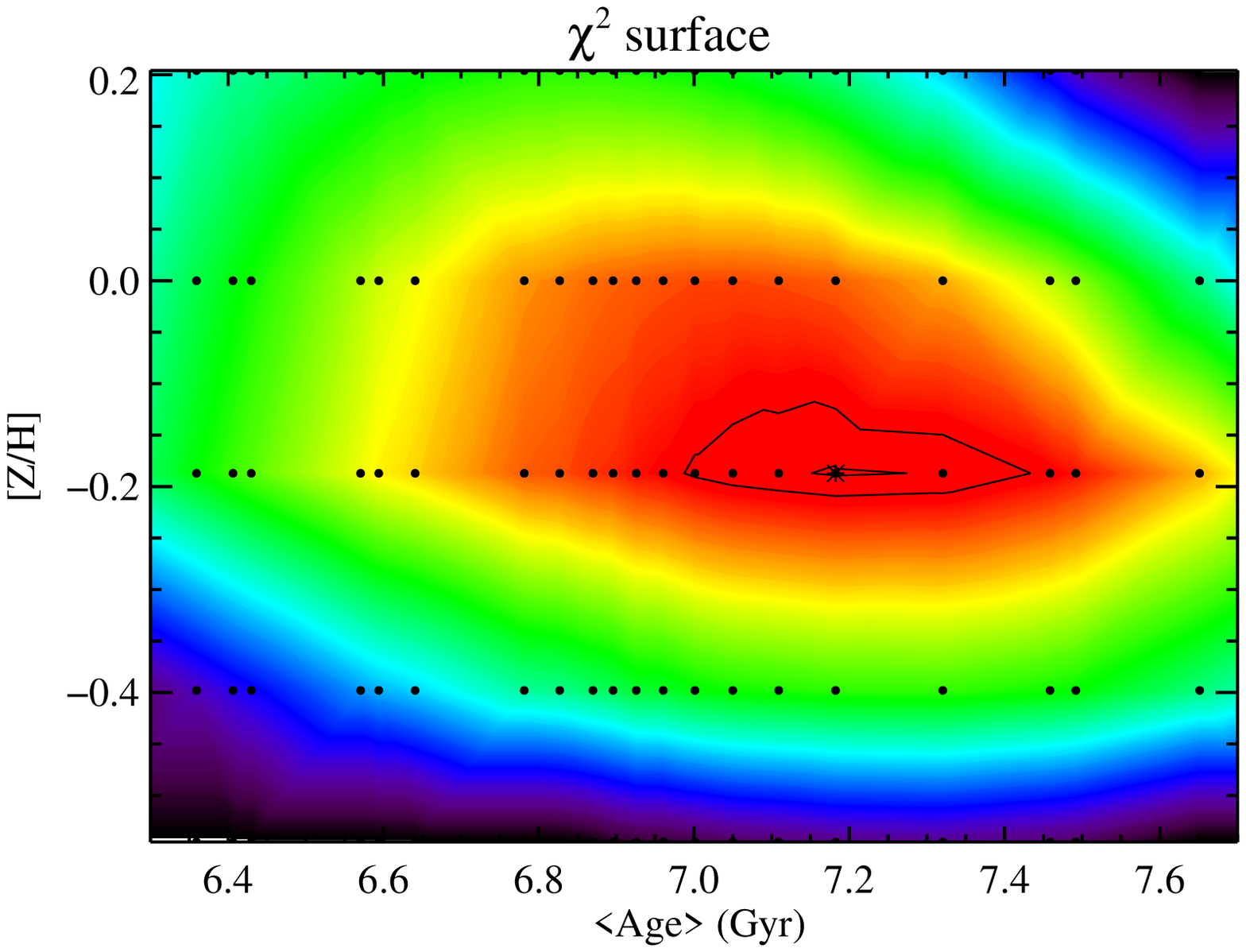}
\plotone{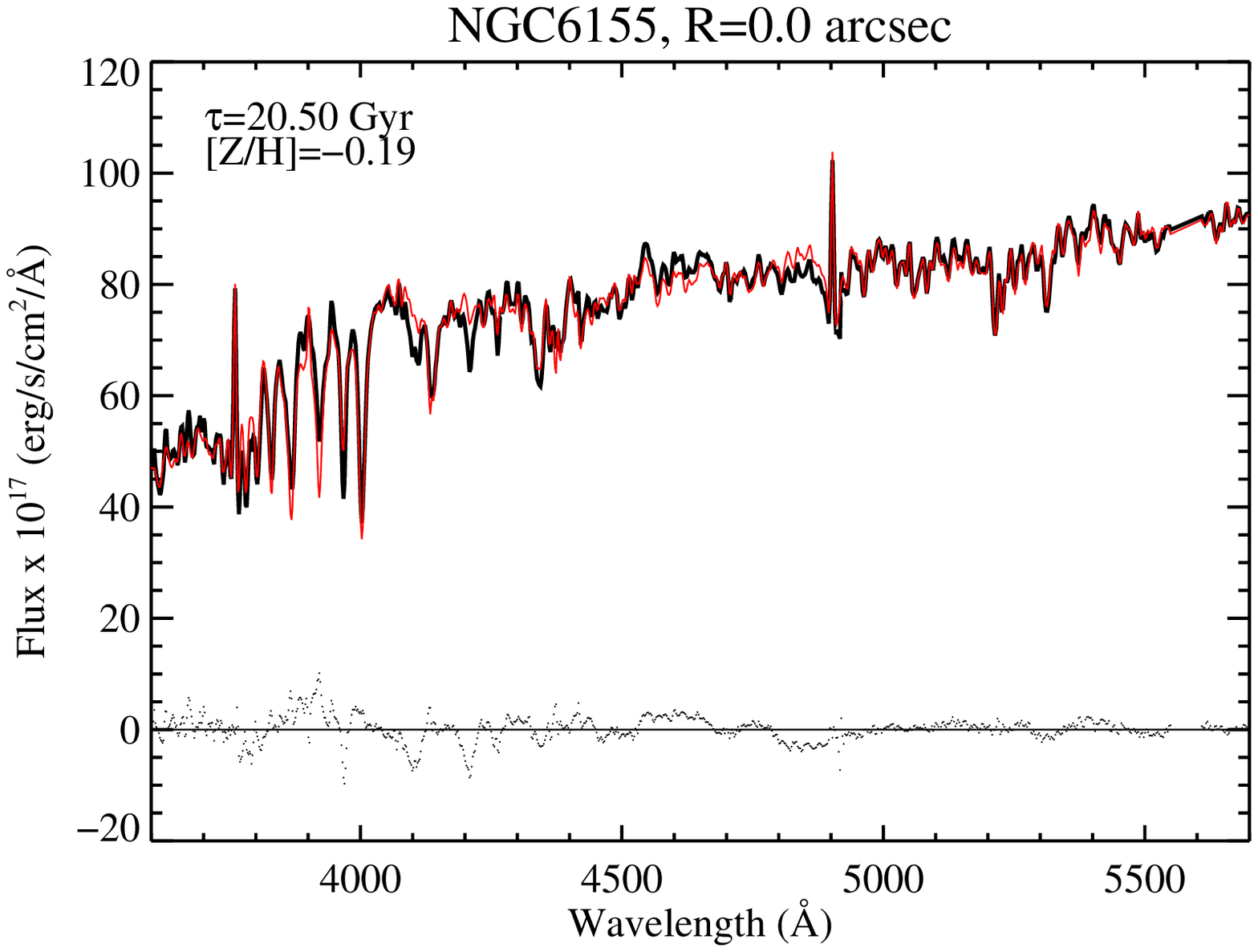}
\caption{An example of our stellar population fitting procedure.  Top:  The computed $\chi^2$ surface.  Each point represents a unique spectral model.  The minimum along with contour levels of $\Delta\chi^2=$20 and 200 are plotted. Bottom:  The observed spectrum (black) with the best-fit model (red).  Residuals are plotted.  \label{spec_fit}}
\end{figure*}

We use the Gas AND Absorption Line Fitting (GANDALF) package \citep{Sarzi06} along with the penalized pixel fitting (pPXF) \citep{Cappellari04} code to find the best-fitting model for each binned spectrum.  The pPXF code fits a stellar velocity and dispersion which is then held constant for the rest of the fit.  GANDALF multiplies model spectra by a low-order polynomial (to correct for dust extinction and flux calibration mis-matches), and simultaneously fits specified emission lines as Gaussians.  Figure~\ref{spec_fit} shows an example of our fitting procedure.  The use of the polynomial ensures that the fit matches the spectral absorption features, and the overall tilt or broadband color does not affect the fit.  Because dust extinction has a negligible effect on the relative depth of absorption features \citep{MacArthur05}, our fitting method does not require correcting the spectra for internal or foreground extinction.

The uncertainty in our fits should be dominated by the systematic difference between the adopted smooth SFH and the bursty SFHs of real galaxies.  To test the errors associated with SFH mismatch, we have generated integrated spectra using the SFHs measured from resolved populations in NGC 300 \citep{Gogarten10} and fit them with our $\tau-$SFHs.  The inner regions of NGC 300 have SFHs strongly peaked at an age of 12 Gyr while the outer regions have constant or increasing SFRs.  The radial bins of NGC 300 have average ages ranging from 5.2-9.6 Gyr.  Fitting our exponential SFHs, we are able to match the relative ages of each radial bin in NGC 300 and have an absolute age dispersion of $\sigma=0.6$ Gyr.  The largest source of error seems to be recent bursts of star formation causing our fitting routine to converge on ages that are younger than the actual mass-weighted stellar population.

Figure~\ref{spec_fit} shows an example of our data and fitting procedure.  The fits become noticeably worse blueward of $\sim$4300 \AA.  There are several possible reasons for the poor fit in the blue.  The instrument throughput in the blue during these observations was low, resulting in poorer SNR.  The templates may need improvement in the blue.  Finally, the poor fit could also be a systematic error left over from imperfect sky subtraction.  We have experimented with  masking this region and find the fits remain basically unchanged.  We emphasize that our errors are clearly non-Gaussian and are dominated by the systematics of template mismatch and SFH mismatch.  Therefore, rather than adopt the standard error bars derived from $\delta\chi^2=2.3$ contour, we plot the larger error bars derived from our experiment fitting the observed SFH of NGC 300.  We also note that a recent burst of star formation could bias our measurements to return values that are too young.

\section{Results}

\begin{figure*}[h]
\epsscale{.6}
\plotone{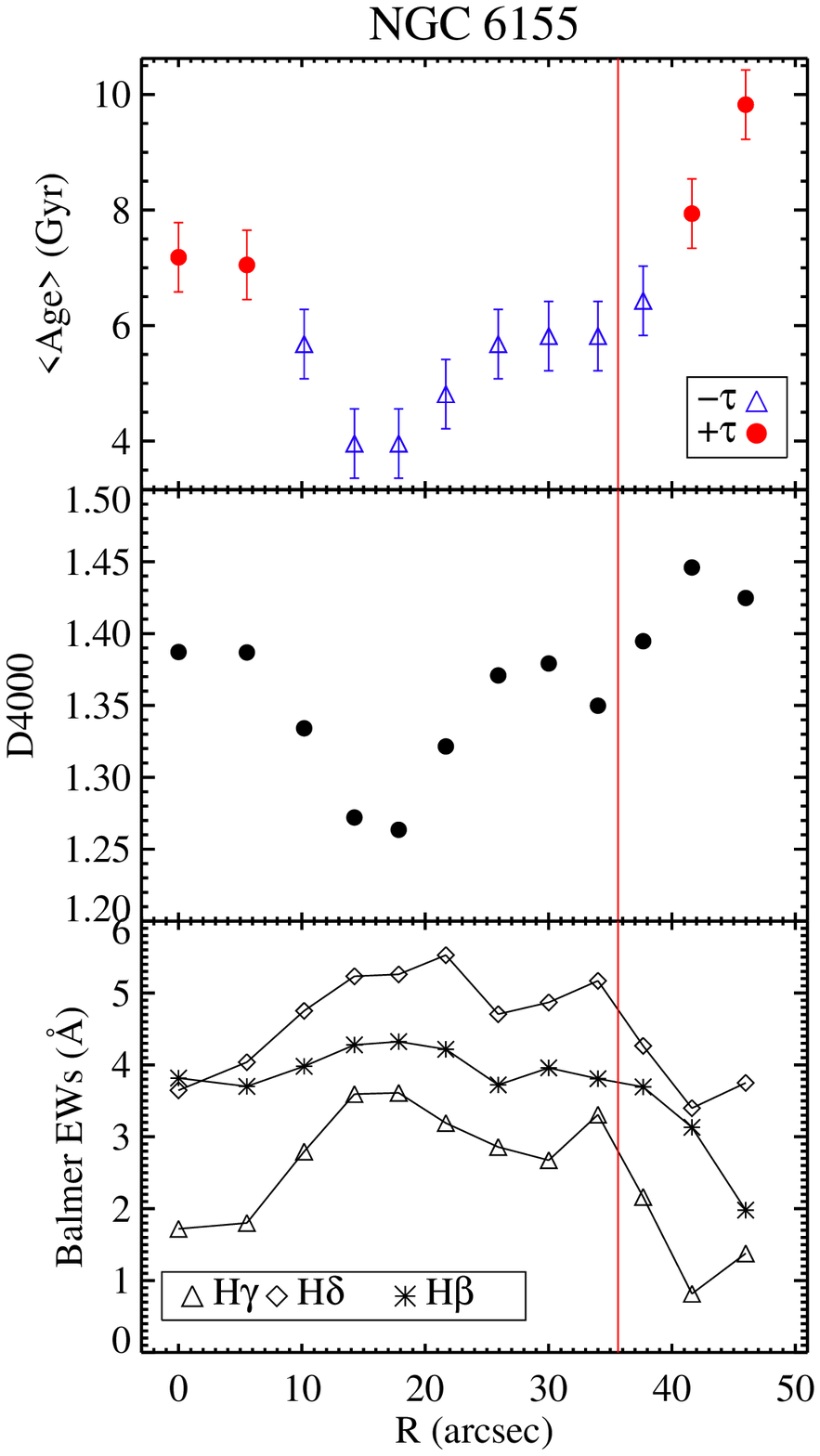}
\caption{{\emph{Top:}} Radial mass-weighted age profile for NGC 6155.  Red circles are used for decreasing star formation histories and blue triangles are used for increasing histories.  The surface brightness break radius is marked with a red vertical line.   Error bars represent our estimated systematic uncertainties.  {\emph{Middle:}} The traditional spectroscopic age indicator D4000.  {\emph{Bottom:}} Emission line corrected Balmer absorption equivalent widths.  We have not matched resolution to formally move onto the Lick system and show the Blamer strengths for illustration.  The D4000 index as well as the Balmer lines all show changes at the profile break consistent with the stellar population becoming older.  The bright star formation ring around R=15\arcsec\ is also clearly measured as younger.  \label{age_plot}}
\end{figure*}

The results of our stellar population fits are plotted in Figure~\ref{age_plot}.  The central region is best fit by a decreasing star formation rate, with a mass-weighted age of $\sim7$ Gyr.  The region at a radius of 15\arcsec\ ($\sim$2.8 kpc) is dominated by very young stars and active star formation.  This is clearly visible in the SDSS image as a region dominated by a star forming ring or inner spiral arms.  Beyond the star forming region, the stars remain young and dominated by recent formation.  Past the surface brightness break radius at 34\arcsec\ ($\sim7$ kpc), there is a sudden change, and the stars are best described with older stellar populations and declining SFHs.  This is an impressive confirmation of the hypothesis that stars beyond the break radius are populated by stellar migration and should show an upturn in stellar age.

Our fits return a metallicity [Z/H]$\sim-0.2$ for most of the region before the profile break.  Beyond the profile break, the metallicity drops to [Z/H]$\sim-0.5$.  While this is consistent with the stellar migration hypothesis, we caution that this is still only a crude measure of the stellar metallicities.

\section{Discussion}

\citet{Pohlen06} describe NGC 6155 as lopsided and classify the galaxy as a Type II-AB (a downbending profile in an asymmetric disk).  While our fiber-photometry shows slight lopsidedness, the velocity field is quite regular and there are no nearby neighbors.  All the galaxies within a 1 Mpc projected radius and similar redshift of NGC 6155 are at least 3 magnitudes fainter.  It therefore seems unlikely that the profile break in NGC 6155 is the result of a dynamical interaction with a neighboring galaxy.

Our result that average stellar ages increase beyond the break radius is consistent with the results of \citet{Bakos08} who stacked multiple SDSS images and found the $g-r$ color became redder beyond the profile break.  This is also similar to the age gradient observed in M33, with the age decreasing out to the profile break where the gradient reverses and the stellar population becomes older beyond the break \citep{Williams09}.  \citet{Bakos08} claim that while their stacked surface brightness profiles shows a break, the stellar mass profile is described by a single exponential.  Our model mass-to-light values are not well constrained beyond the profile break, however, the increase in age we detect does suggest the M/L is larger in this region.  Therefore we expect the stellar mass profile to be smoother than the light profile.

The increase in stellar age we measure is very similar to the simulations of \citet{Roskar08} where the average stellar age jumps from 4 to 6 Gyr.  The best fits beyond the break radius have $\tau\sim0-9.5$ Gyr, consistent with significant ongoing star formation in the region.  We therefore take our results as evidence that while stellar migration can populate outer disks, our fits are consistent with some low-level \emph{in situ} star formation as well.  

Along with the change in stellar age, the surface brightness break also corresponds with a lack of line emission (Figure~\ref{sb}, lower left).  This could be a simple coincidence where the break radius is also where the emission becomes faint enough that we no longer detect it, but it is also consistent with the theory that the break radius represents the location of a star formation threshold.  

It should be pointed out that our observations are also consistent with a radial change in the stellar IMF.  If the outer disk only hosts lower-mass molecular clouds, we could expect star formation to be unable to produce the highest mass stars \citep{Kroupa03,Koeppen06}.  Because young massive stars contribute heavily to the overall luminosity, their absence would result in a young stellar population that spectroscopically appears older.  A change in the IMF can also explain the lack of emission lines beyond the break \citep{Pflamm09}.

In a forthcoming paper, we will present a larger sample of similar observations.  Our larger sample includes galaxies with no surface brightness profile break as well as up-bending profiles.  We also plan to further refine our spectral synthesis template fitting method to improve the metallicity measurements of the stellar populations.

\acknowledgments 

Thanks to L. MacArthur, J. Dalcanton, J. Adams, and G. Blanc for fruitful conversations.  We thank the referee for a thorough review that greatly improved the manuscript.  Thanks to Dave Doss, Earl Green, and the rest of the McDonald Observatory support staff.  We thank M. Sarzi for providing early releases and instruction on the GANDALF code.  Thanks to G. Bruzual and S. Charlot for supplying an early release of their latest models.  PY was supported by the Harlan J. Smith Postdoctoral Fellowship.  This work made use of Craig Markwardt's IDL curve fitting code.



\end{document}